\documentclass[12pt,a4paper]{article}
\usepackage{graphicx}
\usepackage{dcolumn}
\usepackage{bm}
\usepackage{amsmath}
\usepackage{amsfonts}%
\usepackage{amssymb}
\usepackage[version=3]{mhchem} 
\usepackage[mathscr]{eucal}
\usepackage{bbm}
\usepackage[T1]{fontenc} 
\usepackage{pdfpages}
\usepackage{float}       
\usepackage{helvet}      
\usepackage{mathptmx}    
\usepackage{setspace}    
\AtBeginDocument{\doublespacing}

\author{%
   Cheng-Peng Chang \thanks{%
    Center for General Education, Tainan  University of  Technology, 710 Tainan, Taiwan
    E-mail: \texttt{t00252@mail.tut.edu.tw}
   }%
}
\title{Dynamical conductivity of gated AA-stacking multilayer graphene with spin-orbital coupling}

\begin{document}

\maketitle

\begin{abstract}
An efficient method with no numerical diagonalization of a huge Hamiltonian matrix and calculation of a tedious Green's function is proposed to acquire the exact energy spectrum and dynamical conductivity in a gated AA-stacking $N$-layer Graphene (AANLG) with the intrinsic spin-orbital coupling (SOC).  $2N \times 2N$ tight-binding Hamiltonian matrix, velocity operator and   Green's function representation of an AANLG  are simultaneously   reduced to $N$ $2\times 2$ diagonal block matrices through a proper  transformation  matrix. A gated AANLG with intrinsic SOC is reduced to $N$ graphene-like layers. The energy spectrum of a graphene-like layer is  $E= \varepsilon _{\bot}\pm \varepsilon_{||}$. $ \varepsilon _{\bot}$  depends on the interlayer interaction,  gated voltage and layer number.
$ \varepsilon_{||}=\sqrt{E_{MG}^2+ \Delta^2}$, where $E_{MG}$ is the energy spectrum of a monolayer graphene and $ \Delta$ is the magnitude of intrinsic SOC.   More importantly, by inserting the diagonal block velocity operator and Green's function representation in the Kubo formula,  the exact dynamical conductivity of an AANLG  is shown  to be $\sigma = \Sigma_{j=1} ^N  \sigma_j$, the sum of the dynamical conductivity of $N$  graphene-like layers. The analytical form of $\sigma_j$ is presented and the dependence of  $\sigma_j$ on  $\varepsilon_{\bot}$,  $\Delta$, and  chemical potential   is clearly demonstrated. Moreover, the effect of  Rashba SOC on the electronic properties of an AANLG  is  explored with the exact energy spectrum  presented.
\end{abstract}

\section{Introduction}

Graphene and its family members, including the AA-, AB- and ABC-stacking graphenes,  have long  attracted a lot of attention due to their striking physical properties.  Graphene, a pure two-dimensional (2D) system, is  an atomic sheet  peeled off from graphite\cite{Novoselov1,Berger-2004}.   Carbon atoms  are brought together and packed into a hexagonal lattice to form a graphene sheet. Such a geometrical structure consequently  brings about a pair of   low-lying linear energy bands.  Electrons on the graphene sheet behave like  the relativistic massless particles.   This linear dispersion induces a variety of  unique electronic properties, such as, electron-hole symmetry, Klein tunneling,  high mobility at room temperature,  non-zero conductivity, and anomalous quantum Hall  effect \cite{Katsnelson06,Beenakker2008,Jiang2007,Hwang07,Bolotin2008,Du2008,Nair2008,Gomez2007,Katsnelson2006-2,Kim2009,Malard2009,Stauber2008,
Bolotin2009,Balandin2008,Sarma2011}.  Owing to the manifestation of fascinating effects,   graphene is a promising  material  expected to play a vital role  in  technological applications, e.g.,   display screens, electric circuits,  solar cells,  analog electronics and photonics/optoelectronics\cite{Wang2008, WF-2008,Avouris2010, Bonaccorso2010,Dean2010,Schwierz2010,He2012,Avouris2012}.

Multilayer graphenes are  the  pile  of several graphene layers held  together  by the van der Waals force.  The  low-energy physical properties depend strongly both on the stacking order and on  the number of layers\cite{Latil2006,Graf2007,Hass2008,Nilsson2008,Min2008,Lui2010}. The most studied multilayer graphenes are AB-stacking  bilayer graphene\cite{Ohta2006,McCann2006,Ferrari2006,Castro2006,Abergel2007,
Oostinga2007,Zhang2008,Zhang2009,Chang-2014}  and ABC-stacking trilayer graphene\cite{Craciun2009,Koshino2009,Zhang2010,Yuan2011}. The AB-stacking bilayer graphene shows four parabolic bands around the Dirac points. The touch between valence and conduction bands makes  the AB-stacking bilayer graphene  a zero gap semimetal.  A band gap is opened by the application of a vertical electric field\cite{Chang-2014}.  The low energy dispersions of ABC-stacked trilayer graphene  are described  by two remarkably flat bands. The two-fold degeneracy in the band structure can be readily lifted by  a perpendicular electrical field.  Due to the progress in the fabrication and manipulation of graphene layers,  the  AA-stacking graphite and AA-stacking multilayer graphenes are produced\cite{JKLee2008,Borysiuk2011}. Following this,   theoretical and experimental studies  are conducted in order to explore  electronic properties of AA-stacking bilayer and multilayer graphenes, e.g. infrared  spectra, Raman spectra, Landau-level energies, absorption spectra, magneto absorption spectra, static polarization, and dynamical conductivity  \cite{JKLee2008,Borysiuk2011,Liu-2009,YuehuaXu,Chiu2010,Chang2011,MF-Lin2012,Tabert2012,Mohammadi2014}.

The  increase in the  layer number, appearance of  interlayer interaction,   and  application of external field will lead to  more difficulties exploring the electronic properties of the multilayer graphenes.   For instance, a $ 2 N \times 2N$  tight-binding (TB)  Hamiltonian matrix is constructed and used to describe an $N$-layer graphene with the nearest neighbor interactions taken into account.   The  exact diagonalization of  a $ 2 N \times 2N$  Hamiltonian matrix will be cumbersome with the layer number $N$ increasing.  A high-rank  Hamiltonian matrix  gives rise to more tasks in calculation of the Green's function, which is generally adopted to study the minimal and dynamical conductivities.  Moreover,  the mirror-symmetry-breaking, caused by a vertical electric  field  applied to a multilayer graphene,  also increases the difficulty in  the  diagonalization of  Hamiltonian matrix.  Most of studies  focus on the exploration of physical properties as the layer number  $N <3$,  e. g.,  AB-stacking bilayer graphene, and ABC-stacking trilayer graphenes.  Recently, the  investigations  of the dynamical conductivity of AA-stacking  graphene  and  static polarization of AAA-stacking  graphene have been reported \cite{Tabert2012,Mohammadi2014}. A model can deal with physical properties of multilayer graphene in different stacking order  or various layer number $N$ under external field is inspired and desired. We  previously presented analytical modes to exactly describe the minimal conductivity of the AB-stacking multilayer graphene \cite{Chang-2012-1}and exact Landau levels of the AA-stacking multilayer graphene\cite{Chang-2013}.

In this work,  an analytical model is proposed in order to derive the dynamical conductivity and  energy spectrum  in a  gated AANLG with intrinsic SOC.  $2N \times 2N$ Hamiltonian matrix of AANLG is  decomposed   into $N$  $2 \times 2$ diagonal block matrices.  An AANLG is decoupled into $N$   graphene-like layers.   Thus,   a close form of the energy spectrum  is disposed.  Application of  current analytical model to the study of the dynamical conductivity  of AANLG  is conducted. It is shown  that the dynamical conductivity of an AANLG is equal to the sum of the dynamical conductivity of $N$   graphene-like layers with/without intrinsic SOC. Above all,  the presented model can efficiently and exactly give out  the energy spectrum and  dynamical conductivity in a gated AANLG with intrinsic SOC and avoid the diagonalization of  a huge Hamiltonian matrix  and   calculation of associated Green's function.

\section{Gate-Tuned Energy Spectrum  Of AANLG with Spin-Orbital Coupling}

Graphene is a two dimensional atomic sheet made up of carbon atoms, which are precisely packed in a planar hexagonal lattice,  viewed as bipartite lattice composed of two interpenetrating triangular sublattices.   The carbon-carbon bond length   is $b= 1.42{\rm \AA}$ and  the lattice vector is equal to $a=\sqrt{3}b$.  A primitive cell contains two atoms denoted as  $A$ and $B$.  With SOC Taken into consideration,  the Hamiltonian  $H_{MG}$ of a monolayer graphene  is\cite{Kane-2005,Rajyta-2010}
\begin{eqnarray}
H_{MG}=h_0 + h_{ISO} +h_{R},
\end{eqnarray}
where  the first term,  $h_0=\sum_{i,j } \alpha_0  c^+_{i} c_{j} + h. c.$, is  Hamiltonian operator  of the monolayer graphene without SOC.  $c^+_{i} (c_{j})$ is the creation (annihilation) operator and creates (annihilates) an electron at the site $i$  $(j)$. $\alpha_0$ is the intralayer nearest-neighbor hopping  between atoms $A$ and $B$ on the same graphene layer, as illustrated in Fig. 1.  The second term  $h_{ISO}$  is the  intrinsic  spin-orbit interaction.
The third term is the Rashba SOC,  which is induced by the external perpendicular electric field or the interaction with substrate.  The Hamiltonian operator of  Rashba SOC is $h_{R}= -i \lambda_{R}  \sum_{\langle i,j\rangle} \sum_{\mu,\nu} c^+_{i} (\textbf{S}_{\mu,\nu} \times  d_{i,j})_z   c_{j}+ h. c.$, where  $\lambda_{R}$ is the magnitude of the Rashba SOC. $\textbf{S}$ is the Pauli vector,  the subscripts   $\mu$ and $\nu$ represent the spin index, and $d_{i,j }$ is the unit vector pointing from atom site $i$ to its nearest neighbor $j$.

Without the Rashba SOC ($\lambda_{R} =0$), the TB Hamiltonian matrix, spanned by  periodic Bloch functions  $| A\rangle$ and  $| B\rangle$,  is\cite{Tabert2012}
\begin{eqnarray}
H_{MG}=\left( \begin{array}{cc }
\Delta \tau_z s_z    & \alpha_{\bf k}\\
\alpha^*_{\bf k}  & -\Delta \tau_z s_z\\
\end{array}\right),
\end{eqnarray}
where $\alpha_{\bf k}=\alpha_0 f({\bf k})=\alpha_0\sum ^3_{j=1} {\rm exp} (i {\bf k}\cdot {\bf b}_j)$. ${\bf b}_j$ represents the three nearest neighbors on the same graphene plane and ${\bf k}$ is the in-plane wave vector. $\Delta$ is the strength of  ISOC and $s_z= \pm 1$ represents the up or down spin.  $ \tau_z=\pm 1$  at the Dirac points K and K'.  The energy dispersions are $E=  \pm \sqrt{ |\alpha_{\bf k}|^2 +\Delta ^2 }$.


Furthermore,  the TB Hamiltonian matrix of the Hamiltonian $H_{MG}=h_0 + h_{ISO} +h_{R}$, acting on  periodic Bloch functions  $| A \uparrow \rangle,    | B \uparrow \rangle, | A \downarrow \rangle,   | B \downarrow \rangle$,  is\cite{Rajyta-2010}
\begin{eqnarray}
H_{MG}=\left( \begin{array}{cccc}
   \Delta                 & \alpha_{\bf k}      & 0                         & 0                \\
\alpha^*_{\bf k}  &   -\Delta                & -i\lambda_R       &0        \\
  0                         &   i\lambda_R         &  -\Delta               & \alpha_{\bf k} \\
 0                          &  0                            &\alpha^*_{\bf k}  & \Delta \\
\end{array}\right),
\label{Hamil-MG}
\end{eqnarray}
where  the Rashba SO  interaction between $| A \uparrow \rangle$ and  $ | B \downarrow\rangle$ ($ | B \uparrow\rangle$ and   $| A \downarrow \rangle$)   is neglected because it is much weaker than $\lambda_R$\cite{Rajyta-2010}. The analytical energy dispersions are
\begin{eqnarray}
\Lambda_{\pm\pm}=  \pm \frac{\lambda_R}{2} \pm \sqrt{ |\alpha_{\bf k}|^2 +\Delta ^2+\lambda_R  \Delta   +\frac{\lambda^2_R}{4} }.
\label{Energy-MG}
\end{eqnarray}


\subsection{Energy Spectrum  Of AANLG with Intrinsic SOC}

By stacking  $N$ layer graphenes directly on each other with an interlayer distance between graphenes $c=3.35$ {\rm \AA}\cite{Charlier1}, an AANLG is formed, as shown in Fig. 1.  In the stacking direction,  $N$ atoms $A$ ($B$) form a linear chain.  The primitive   unit cell contains $2N$ atoms, denoted as $A_1,A_2, \cdots   A_N, B_1, B_2, \cdots, B_N$,.    The first Brillouin zone is the same as that of a graphene.  The Hamiltonian  of an AANLG is given by
\begin{eqnarray}
H_{AANLA} =H_0+ H_{ISO} +H_{R},
\end{eqnarray}
where $H_0$ is Hamiltonian operator an AANLG  without SOC, and  $H_{ISO}$ and $H_{R}$ are caused by ISO and Rashba SO interactions.  Following the discussion in the subsection above, we first close the Rashba effect, i. e. ,   $H_{R}=0$. In the presence of an electric field, the Hamiltonian representation of an AANLG with intrinsic SOC, spanned by  periodic Bloch functions $|A_{1}\uparrow\rangle , |A_{2}\uparrow\rangle, \cdots   |A_{N}\uparrow\rangle$, $|B_{1}\uparrow\rangle , |B_{2}\uparrow\rangle, \cdots   |B_{N}\uparrow\rangle$,   is a $ 2N \times 2N$ matrix,  reading
\begin{subequations}
\begin{eqnarray}
H=\left( \begin{array}{cc }
H_{AA}    & H_{AB}\\
H_{BA}    & H_{BB}\\
\end{array}\right),
\label{Hamil-0}
\end{eqnarray}
where $H_{AA},  H_{AB}, H_{BA}$,  and   $H_{BB}$ are $N\times N$ matrices.   The Hamiltonian operators  $H_{AA}$ and $ H_{AB}$  describe  an  $N$-site linear chain  with the intrinsic SOC subjected to a parallel electric field.

The recent researches\cite{Konschuh-2012, Kormanyos-2013} , using ab intial calculation and TB method,  show that in the Bernal stacking  bilayer graphene and  ABC stacking trilayer graphene, many  intralayer and  interlayer  SOC parameters  are included  to fully  describe the spin-orbital interactions in the TB model.
The numerical fitting of TB parameters to the results of  ab initial calculation exhibits that
the magnitudes of intrinsic ($\Delta$) and Rashba spin-orbital interactions ($\lambda_R$) are  layer-position-dependent. What's more, the strength of the interlayer intrinsic SOC is much weaker than  that  of the  intralayer  intrinsic SOC. Accordingly, we neglect the  interlayer intrinsic spin-orbital interactions in our case and then  take  the  intralayer  intrinsic  and Rashba spin-orbital interactions into account. Thus, the  two $4 \times\ 4$ $H_{AA}$ and  $H_{BB}$ Hamiltonian matrices of  the AA-stacking quad-layer graphene, for instance, are expressed as the following,
\begin{eqnarray}
H_{AA}=\left( \begin{array}{cccc }
 3V/2 + \Delta_1    &\alpha_1 & 0      & 0  \\
\alpha_1& V/2+ \Delta_2 & \alpha_1  & 0  \\
  0      &\alpha_1 & -V/2  + \Delta_3& \alpha_1 \\
  0      & 0             &\alpha_1 & -3V/2  + \Delta_4  \\
\end{array}\right),
\end{eqnarray}
\begin{eqnarray}
H_{BB}=\left( \begin{array}{cccc }
3V/2 - \Delta_1    &\alpha_1            & 0              & 0\\
\alpha_1          & V/2- \Delta_2        & \alpha_1  & 0  \\
0     &\alpha_1 & -V/2- \Delta_3      & \alpha_1  \\
0      & 0             &\alpha_1      & -3V/2- \Delta_4\\
\end{array}\right),
\end{eqnarray}
 where  $\Delta_j$ ($j=1,2,3$ and 4) is the intralayer  intrinsic spin-orbital interaction of the $j$-layer graphene.
 $V= |e|F c$ is the effect electric potential difference between the adjacent layers caused by the external electric field.
Then, $V$ is denoted as the gate voltage. In  multilayer graphenes, the  potential drop between two adjacent layers  might be affected by the screening  processes\cite{Gelderen-2013}. For simplicity,  we also assume that the potential drop is the same on each  graphene layer.  The interlayer hopping parameter, $\alpha_1$, couples the two $A$ (or $B$) atoms from the two adjacent layers [Fig. 1].  $\alpha_3$, the interlayer interaction between atoms $A$ and $B$ from the two adjacent layers,  results in a weak electron-hole asymmetry in an AANLG \cite{Chang-2013,Charlier1}. The values of the hopping integrals are $\alpha_1=0.361$ eV and  $\alpha_3=-0.032$ eV \cite{Charlier1}.  Only the  main interlayer interaction $\alpha_1$ is taken into consideration because of  $\alpha_3 \ll\alpha_1$.
  The matrix element $H_{AB}$,  resulting from the intralayer interaction, reads
\begin{eqnarray}
H_{AB}=H_{AB}^*=\left( \begin{array}{cccc}
 \alpha_{\bf k}  & 0   & 0   & 0\\
 0    & \alpha_{\bf k}  & 0  & 0\\
 0   &0    &\alpha_{\bf k}   &0  \\
 0   &0   &0      &\alpha_{\bf k}   \\
\end{array}\right).
\end{eqnarray}
\end{subequations}

The layer-position-dependent intralayer  intrinsic spin-orbital interaction   and Rashba  SOC destroy the inversion symmetry of AANGL.  The breaking  of the inversion symmetry  complicates the analysis and discussion. For simplicity,  the intralayer  intrinsic spin-orbital interaction  are assumed to be independent of the vertical positions; that is,   $\Delta_j=\Delta$ and $\lambda_{R,j}=\lambda_{R}$.  Notably,  the  $H_{AA}$ ($H_{BB}$)  is the sum of two matrices $H_V$ and $H_{ISO}$ and reads
\begin{subequations}
\begin{eqnarray}
H_{AA}= H_V + H_{ISO}=  H_V + \Delta \mathbbm{1},
\label{H-AA}
\end{eqnarray}
\begin{eqnarray}
H_{BB}= H_V + H_{ISO} = H_V - \Delta \mathbbm{1},
\label{H-BB}
\end{eqnarray}
\end{subequations}
where $\mathbbm{1}$ is a $N \times N$ identity matrix.  $H_V$ describes  an  $N$-site linear chain  without SOC subjected to a parallel electric field.  $H_{AA}$ ($H_{BB}$)   is commute with $ H_V$; that is,   $H_{AA}$ ($H_{BB}$)   and $ H_V$ share the same eigenfunctions.   The eigenenergy $\varepsilon_j$ and associated   eigenfunction $ |\textsf{S}_j \rangle$ of  $H_V $are easily obtained through the diagonalization of the  eigenvalue equation\cite{Chang-2013}
\begin{eqnarray}
H_V |\textsf{S}_j \rangle = \varepsilon_j |\textsf{S}_j \rangle,
\label{Eigen-EQ}
\end{eqnarray}
where $j=1, 2,  \cdots, N$.   The  transpose of   $|\textsf{S}_j\rangle $  is   $|\textsf{S}_j \rangle^T=|s_{j,1}$,  $s_{j,2}$,  $s_{j,3}$,  $\cdots$ $s_{j,N}\rangle$ and the component  $s_{j,l}$, is the site amplitude of atom $A$ or $B$ located at the $l$th layer.

With  column vectors  $ |\textsf{S}_j\rangle  $,  the eigenfunctions,    an $ N \times  N$ unitary transformation matrix $\hat U_V=(|\textsf{S}_1 \rangle, |\textsf{S}_2  \rangle, \cdots , |\textsf{S}_N \rangle )$ is then constructed  and used to diagonalize  $H_V$, i. e.,    $\hat U_V^\dag H_V \hat U_V=   \varepsilon_j   \mathbbm{1}$, where $ \mathbbm{1}$ is a unit matrix.  The eigen-energies of   $H_{AA}$ and $H_{BB}$ are  $\varepsilon_j + \Delta$ and $\varepsilon_j - \Delta$ after the diagonalization of Eqs. (\ref{H-AA}) and  (\ref{H-BB}), respectively .

To acquire the energy spectrum of  AANLG,  a $ 2N \times 2N$  unitary transformation  matrix
\begin{eqnarray}
\textbf{U}=\left( \begin{array}{cc }
\hat U_V    &  0 \\
0    &  \hat U_V\\
\end{array}\right),
\label{Utrans}
\end{eqnarray} is built to transform the Hamiltonian matrix (Eq. (\ref{Hamil-0}) ) into a simple form.
After the unitary transformation,   a reduced matrix  $\mathcal{H}_{red}=\textbf{U}^\dag  H \textbf{U} $  has the form
\begin{eqnarray}
\mathcal{H}_{red}=\left( \begin{array}{cc }
  (\varepsilon_j +\Delta ) \mathbbm{1}   &  \alpha_{\bf  k} \mathbbm{1}\\
  \alpha^*_{\bf  k} \mathbbm{1}   &   (\varepsilon_j  -\Delta ) \mathbbm{1}     \\
\end{array}\right),
\label{Utrans}
\end{eqnarray}
 where $\mathbbm{1}$ is $N\times N$ unit matrix. Then,   the reduced Hamiltonian matrix can be  rearranged  into block diagonal form, $\mathcal{H}_{red} = \mathcal{H}_{1}\oplus \mathcal{H}_{2}\oplus \cdots \oplus \mathcal{H}_{N}$, where each  $2 \times 2$ block diagonal matrix $\mathcal{H}_j$ is expressed as follows
\begin{eqnarray}
\mathcal{H}_j=
\left( \begin{array}{cc}
\varepsilon_j +\Delta   &   \alpha_{\bf  k}\cr
 \alpha^*_{\bf  k}    &\varepsilon_j -\Delta\\
\end{array} \right).
\label{subsys}
\end{eqnarray}
That is to say, an AANLG  can be decomposed into $N$ subsystems,  $\mathcal{H}_j$.  The exact  energy spectrum of each  subsystem is $$E_{j, \pm} = \varepsilon_j   \pm  \sqrt{|\alpha_{\bf  k}|^2 + \Delta^2   }= \varepsilon _{\bot} \pm  \varepsilon_{||},$$   where $\varepsilon_j ( =\varepsilon _{\bot} )$ depends on the magnitude of interlayer interaction,  gated voltage and layer number.  $ \varepsilon_{||}=  \sqrt{|\alpha_{\bf  k}|^2 + \Delta^2  }$ is the energy spectrum of a monolayer graphene with SOC.

 Around the Dirac point $K$, the diagonal block is for $\textbf{k = K + q}$
\begin{eqnarray}
\mathcal{H}_j=
\left( \begin{array}{cc}
\varepsilon_j  +\Delta &   -\hbar v_F (q_x + iq_y)\cr
-\hbar v_F (q_x - iq_y)    &\varepsilon_j -\Delta\\
\end{array} \right),
\label{subsys-K}
\end{eqnarray}
and $\hbar v_F= \frac{3}{2}\alpha_0 b$ is the Fermi velocity. The low-lying energy dispersions associated with $\mathcal{H}_j$ are $ E_{j, \pm} = \varepsilon_j   \pm  \sqrt{ | \hbar v_F q|^2   + \Delta^2   }$,  where $|q|=\sqrt{q_x^2+q_y^2}$.

\subsection{Energy Spectrum  Of AANLG with Intrinsic and Rashba  spin-orbital Interactions}

If we take  both the intrinsic and Rashba  spin-orbital interactions into consideration, TB Hamiltonian matrix of an AANLG subject to a perpendicular electric field, acting on  periodic Bloch functions
$| A_1 \uparrow \rangle,    | B_1 \uparrow \rangle, | A _1\downarrow \rangle,   | B_1 \downarrow \rangle$,
  $\cdots$, $| A_j \uparrow \rangle,    | B_j \uparrow \rangle, | A _j\downarrow \rangle,   | B_j \downarrow \rangle$,  $ \cdots$, $| A_N \uparrow \rangle,    | B_N \uparrow \rangle, | A _N\downarrow \rangle,   | B_N \downarrow \rangle$, is a $ 4N  \times 4N$ Hermitian matrix and expressed as follows
\begin{eqnarray}
H_{AANLG}=
\left( \begin{array}{cccccc}
H_1&H_T&0      &\cdots&\cdots & 0 \cr
H_T&H_2&H_T& \ddots         &\vdots &0\cr
    0 &H_T&H_3 &H_T&0&\cdots     \cr
0&0& \ddots   & \ddots   &\ddots  &\vdots  \cr
0  &\vdots  &\ddots &\ddots &\ddots&H_T\cr
0 & \cdots &0 &\cdots  & H_T  &H_N\cr
\end{array} \right),
\end{eqnarray}
where $ H_j$ and $H_T$ are $4 \times 4$ blocks.  The off-diagonal block   $H_T=  \alpha_1  \mathbbm{1}$ originates in the  main interlayer interaction $\alpha_1$.  The diagonal block $H_j=   V_j  \mathbbm{1} +H_{MG}$ is the Hamiltonian matrix of the $j$-layer graphene in the presence of the gated potential $V_J$, which has the form
\begin{eqnarray}
H=\left( \begin{array}{cccc}
  V_j+ \Delta                 & \alpha_{\bf k}      & 0                         & 0                \\
\alpha^*_{\bf k}  &   V_J-\Delta                & -i\lambda_R       &0        \\
  0                         &   i\lambda_R         &  V_J-\Delta               & \alpha_{\bf k} \\
 0                          &  0                            &\alpha^*_{\bf k}  & V_J+\Delta \\
\end{array}\right).
\end{eqnarray}
It is easy to diagonalize the block $H_j=   V_j  \mathbbm{1} +H_{MG}$ through a  $ 4 \times 4$ unitary transformation matrix $\mathbbm{U}$, which transforms  $H_{MG}$  into a diagonal matrix, .i. e.,   $\mathbbm{U}^+ H_{MG}\mathbbm{U}= diag (\Lambda_{++},\Lambda_{+-},\Lambda_{--},\Lambda_{-+})$ (Eq. (\ref{Energy-MG})).  The eigenvalues of $H_j$ are $\Lambda_{J,\pm\pm}=V_J +\Lambda_{\pm\pm}$. Then,   a $ 4 N \times 4N$ unitary transformation matrix  $\textbf{U}= diag(\mathbbm{U} , \mathbbm{U},\dots, \mathbbm{U})$ is constructed and used to transform $H_{AANLG}$ into a diagonal block form. After the operation, we  obtain  $ \textbf{U}^+  H_{AANLG}  \textbf{U}= H_{++} \oplus H_{+-} \oplus H_{- -} \oplus H_{-+} $,  where $\mathcal{H}_{\eta\xi}$ ( $\eta=\pm, \xi=\pm$)   is an $N\times N$ matrix.

We take  the AA-stacking trilayer graphene as a study model. The Hamiltonian matrix $H_{AATLG}$ and unitary transform  matrix $\textbf{U}$ are
\begin{eqnarray}
H_{AATLG}=
\left( \begin{array}{ccc}
H_1&H_T&0 \\
H_T&H_2&H_T\\
    0 &H_T&H_3 \cr
\end{array} \right),
{~~~\rm  and~~~~}
U=
\left( \begin{array}{ccc}
\mathbbm{U}&0&0 \\
0&\mathbbm{U}&0\\
0 &0&\mathbbm{U} \cr
\end{array} \right).
\end{eqnarray}
After the  operation $ \textbf{U}^+  H_{AATLG}  \textbf{U}$,    $H_{AATLG}$ is arranged into the block diagonal form    $H_{AATLG}=H_{++} \oplus H_{+-} \oplus H_{- -} \oplus H_{-+}$, where each   $H_{\eta\xi}$  is $3 \times 3$ matrix and $H_{\eta\xi}= \Lambda _{\eta\xi} \mathbbm{1} + H_V$.  The latter term   $H_V$,  Eq. (\ref{H-AA}),  describes  an  $N(=3)$-site linear chain  without SOC subjected to a parallel electric field.   The energy spectrum related to  $H_{\eta\xi}$ are $E= \Lambda _{\eta\xi}+\varepsilon_j = \varepsilon_{||}+\varepsilon_{\perp}$, where $ \Lambda _{\eta\xi} = \varepsilon_{||}$  and  $\varepsilon_{\perp} = \sqrt{V^2 +2\alpha^2_1}, ~~0,~~ {\rm  or}~~-\sqrt{V^2 +2\alpha^2_1}$\cite{Chang-2013}.

\section{Electronic Properties and Discussions}

The energy dispersions  of  an AANLG with SOC in the presence of  the gated potential are easily  obtained through the calculation of  energy spectrum of each  subsystem by using  analytical formula $E_j= \varepsilon_{\perp}+\varepsilon_{||} $.  For example, the energy spectrum  of the AA-stacking bilayer graphene  are described as   $E=\pm \sqrt{\alpha_1^2 +v^2/4}+  \Lambda _{\pm\pm}$, where $ \Lambda _{\pm\pm}=  \pm  \frac{\lambda_R}{2} \pm \sqrt{ |\alpha_{\bf k}|^2 +\Delta ^2+\lambda_R  \Delta   +\frac{\lambda^2_R}{4}}$. Since the energy dispersions are symmetry about $E=0$, only the energy spectrum $ E  >  0 $ are shown in Fig. 2.   In the absence of the gates potential and SOC, the energy dispersions around the Dirac  point $\textbf{K}$  illustrate one pair of linear bands crossing at $E=\alpha_1$ (dashed curves in the inset). The gated potential $V=0.4 \alpha_1$ shifts the linear bands upward (red solid curves  in the inset).  In the AB-stacking bilayer graphene, the intrinsic SOC parameter is $\Delta \sim 10^{-5}$  eV and hence $\Delta/  \alpha_1 \sim 10^{-4}$\cite{Konschuh-2012}.   For convenience of  numerical analysis, we use $\Delta/  \alpha_1=0.1$ in this work. The analytical model and numerical results are relevant and  applicable to the  exploration of  physical properties in  multilayer graphene-like systems.  The inclusion of  the intrinsic SOC $\Delta= 0.1 \alpha_1$  changes the linear bands  (red solid curves) into the parabolic bands (blue solid curves), which are described by $E=\sqrt{\alpha_1^2 +v^2/4} + \sqrt{ | \hbar v_F q|^2   + \Delta^2} $.  The  maximum (minimum) of the parabolic band, located at the Dirac point $K$,  is  $E=(\sqrt{1.16 } -0.1)\alpha_1 $ ($E=(\sqrt{1.16 } + 0.1)\alpha_1 $).  The green curves  are the energy spectrum of the gated AA-stacking bilayer graphene with  the Rashba SO interaction  $\lambda_R =0.05 \alpha_1$. The Rashba SO interaction $\lambda_R$ destroys the degeneracy of the linear bands and produces  four parabolic bands (green curves). The middle two parabolic bands  touch each other at $E=\sqrt{1.16 }\alpha_1$.

The energy dispersions of the AA-stacking trilayer graphene are  also evaluated with   formula $E= \Lambda _{\pm \pm}+\varepsilon_j$, where   $\varepsilon_{1} = \sqrt{V^2 +2\alpha^2_1}$, $\varepsilon_{2}= 0$, and $\varepsilon_{3}=- \sqrt{V^2 +2\alpha^2_1}$.  Without SOC, the  energy spectrum $E_2= \Lambda _{\pm \pm}+\varepsilon_2$ are independent of the magnitude  of  gated potential, as shown by the black dashed and red solid curves in the inset of Fig. 3.  Two linear bands cross over at $E=0$.    The   intrinsic SOC  $\Delta$ changes the linear  into the parabolic bands, as  illustrated by dashed blue curves, which is simulated by  $E_2= \sqrt{ | \hbar v_F q|^2   + \Delta^2} $, The  maximum (minimum) of the parabolic band is determined by the strength of   $\Delta= 0.1  \alpha_1$.  There are four parabolic bands after the inclusion of the  Rashba SOC  (green cures). The middle two parabolic bands do not touch at $E=0$  due to the intrinsic SOC.

\section{Green's function and Velocity operator}

After the introduction of the Rashba effect, as illustrated in the section above, the $ 4 N \times 4N$ Hamiltonian matrix  can be divided into four  $  N \times N $ diagonal blocks.   This would complicate the discussion and block us to pursue a simple analytical form of the conductivity of an AANLG.  We, then, switch off the Rashba effect and consider the intrinsic SOC (ISOC)  alone in the following work. Now, we  exhibit that the Green's function and   velocity  operator associated with an AANLG  can be transformed into  the diagonal block matrices. With the  Hamiltonian matrix $H$,  it is straightforward  to calculate the Green's function  through  $G(z) = \frac{1}{zI - H} $.  The larger Hamiltonian matrix gives rise to  more complex tasks  in calculation of  the inverse matrix of ${zI - H}$.  To reduce the task, we use the unitary operator $\textbf{U}$, which  causes  $\mathcal{H}_{red}=\textbf{U}^\dag  H \textbf{U}$,  to transform the Green's function. After the operation, we have $\mathcal{G}=\textbf{U}^\dag  G\textbf{U} = \textbf{U}^\dag  \frac{1}{zI - H}  \textbf{U}$  $= \frac{1}{zI -  \mathcal{H}_{red}}  $.  $\mathcal{H}_{red}$ is a block diagonal matrix and so does $(zI - \mathcal{H}_{red})$. Now, the  Green's function $\mathcal{G}$ is also in  a block diagonal form; that is,  $\mathcal{G} = \mathcal{G}_{1}\oplus \mathcal{G}_{2}\oplus \cdots \oplus \mathcal{G}_{N}$.  Moreover,  each sub-Green's function is $\mathcal{G}_{j}= \frac{1}{zI -  \mathcal{H}_{j}}$ and it is a $ 2 \times 2$ matrix,
\begin{eqnarray}
\mathcal{G}^{-1}_{j}=
\left( \begin{array}{cc}
z-\varepsilon_j -\Delta   &  - \alpha_{\bf  k}\cr
 -\alpha^*_{\bf  k}    &z-\varepsilon_j+\Delta\\
\end{array} \right),
\label{subsys-G}
\end{eqnarray}
with the corresponding elements
\begin{eqnarray}
&\mathcal{G}_{j,11}(z)=   \frac{z-\varepsilon_j+\Delta}{(z-\varepsilon_j)^2 - (\Delta^2 + |\alpha_{\bf  k}|^2) }
                                  & =     \frac{ \frac{1}{2}{\Big[} 1 + \frac{\Delta} {\sqrt{ \Delta^2 +|\alpha_{\bf  k}|^2} } \Big] }   { z- E_{j,+} }
                                        + \frac{ \frac{1}{2}\Big[ 1 - \frac{\Delta} {\sqrt{ \Delta^2 +|\alpha_{\bf  k}|^2} } \Big] }   {  z-E_{j,-}  }  \cr
&\mathcal{G}_{j,12}(z)=   \frac{ \alpha^*_{\bf  k} }   {(z-\varepsilon_j)^2 - (\Delta^2 + |\alpha_{\bf  k}|^2) }
       & =    \frac{\alpha^*_{\bf  k}} {2 \sqrt{ \Delta^2 +|\alpha_{\bf  k}|^2} }   \Big[ \frac{1  }   { z- E_{j,+} }
                                      -  \frac{ 1 } { z- E_{j,- } } \Big ]  \cr
&\mathcal{G}_{j,21}(z)  =    \frac{ \alpha_{\bf  k} }       {(z-\varepsilon_j)^2 - (\Delta^2 + |\alpha_{\bf  k}|^2) }
                   & =    \frac{\alpha_{\bf  k}} {2 \sqrt{ \Delta^2 +|\alpha_{\bf  k}|^2} }   \Big [ \frac{1  }   { z- E_{j,+} }
                                      -  \frac{ 1 } { z- E_{j,- } }\Big ] \cr
&\mathcal{G}_{j,22}(z)=   \frac{z-\varepsilon_j -\Delta}{(z-\varepsilon_j)^2 - (\Delta^2 + |\alpha_{\bf  k}|^2) }
                                    & =     \frac{ \frac{1}{2}\Big [ 1 - \frac{\Delta} {\sqrt{ \Delta^2 +|\alpha_{\bf  k}|^2} } \Big] }   { z- E_{j,+} }
                                      + \frac{ \frac{1}{2}\Big[ 1 + \frac{\Delta} {\sqrt{ \Delta^2 +|\alpha_{\bf  k}|^2} } \Big ]}   {  z-E_{j,-}  }.
\label{G-elements}
\end{eqnarray}

The velocity operator,  $\textbf{V} = \frac{\partial H}{ \partial \hbar\textbf{k}}$, is  approximated as  the derivative of the Hamiltonian with respect to the momentum  $\hbar\textbf{k}$,  based on the gradient approximation. According to  the Eq. (\ref{Hamil-0}), the velocity matrix related to AANLG is
\begin{eqnarray}
\textbf{V}=\left( \begin{array}{cc }
\frac{\partial   H_{AA} }{ \partial \textbf{k}}   & \frac{\partial   H_{AB}}{\partial \textbf{k}}\\
\frac{\partial  H_{BA}}{\partial \textbf{k}}    &  \frac{\partial H_{BB} }{ \partial \textbf{k}}\\
\end{array}\right).
\label{Velocity-0}
\end{eqnarray}
Since $H_{AA}$ and $H_{BB}$ are independent of the wave vector $\textbf{k}$,   $\frac{\partial   H_{AA} }{ \partial \textbf{k}} $ and $ \frac{\partial H_{BB} }{ \partial \textbf{k}}$ are equal to zero. As a result,  the velocity matrix is
 \begin{eqnarray}
 \textbf{V}_x =\left( \begin{array}{cc }
    0          &   v_f \mathbbm{1}  \\
v_f\mathbbm{1}  &        0   \\
\end{array}\right),
\end{eqnarray}
here $\mathbbm{1}$ is an $N \times N$ identical matrix and $v_f=\frac{3 b\alpha_0}{2\hbar }$ is the Fermi velocity.  After the action of the  transformation matrix,  $\mathcal{V} = U^\dag V_x U $,  the transferred velocity matrix is a diagonal block matrix, which is in the form of $\mathcal{V} = \mathcal{V} _{1}\oplus\mathcal{V} _{2}\oplus \cdots \oplus\mathcal{V} _{N}$.   Each  $\mathcal{V} _{j}$, a 2 $\times 2$ matrix, is
\begin{eqnarray}
\mathcal{V} _{j}=\left( \begin{array}{cc }
    0          &   v_{f} \\
v_{f}    &        0   \\
\end{array}\right).
\end{eqnarray}
The unitary transformation matrix $\textbf{U}$,    diagonal block  Green's function  representation  $\mathcal{G}$, and velocity operator $\mathcal{V}$ are now utilized to derive the analytical  form of the dynamical conductivity of AANLG.

\section{Dynamical Conductivity of AANLG}

The finite frequency conductivity is studied by  using the Kubo formula. The conductivity is written in terms of  the imaginary part of retarded current-current correlation function divided by  frequency  $\Omega$ as 	 $\sigma_{\alpha \beta}= \frac{\rm{Im}~\Pi_{\alpha  \beta}(\Omega+i0^{\dagger})} {\Omega}$,  where  $\Pi_{\alpha  \beta}(\Omega)$ is also referred to as the polarization function.  Furthermore, the polarization function can then be written in the  bubble approximation as
\begin{eqnarray}
\Pi_{\alpha \beta}(i \nu_m)=
e^2 T \sum_{i\omega_n}\int \frac{d^2 k}{(2\pi)^2} {\rm Tr} [V_{\alpha}~ G(i\omega_n +i\mu_m, {\bf k})
~ V_{\beta}~ G(i\omega_n,{\bf k})],
\end{eqnarray}
where $V_{\alpha}$ is the velocity operator in the direction $\alpha = x$ or $y$ and  $G(i\omega_n,{\bf k})$ is the  Green's function.   With the spectral function representation
\begin{eqnarray}
G_{nm}(z)= \int^{\infty}_{-\infty} \frac{d \omega}{2\pi}
\frac{A_{nm}(\omega')}{z-\omega'},
\end{eqnarray}
the real part of the conductivity, at the zero temperature $T=0$,  is expressed as
\begin{eqnarray}
\sigma_{\alpha \beta}(\Omega)=
\frac{e^2}{2\Omega}\int^{\infty}_{-\infty}\frac{d \omega}{2\pi}
[f(\omega-\mu)-f(\omega+\Omega-\mu)]\cr
\times \int \frac{d^2 k}{(2\pi)^2}
{\rm Tr} [V_{\alpha}~ A(\omega+\Omega,{\bf k})
~ V_{\beta}~  A(\omega,{\bf k})],
\label{conductivity-2}
\end{eqnarray}
where $f(x)=1/[{\rm exp}(x/T) +1]$ is the Fermi function and $\mu$ is the chemical potential.
Following the  aforementioned  method, the AC conductivity for AANLG can be directly calculated  by putting the $2N \times 2N$   Green's function representation (or spectral function representation)  and $2N \times 2N$  velocity  operator in Eq. (\ref{conductivity-2}). The larger the  Green's function (or spectral function representation)  is, the  more calculation tasks are.

To make less complex, we first utilize the relation, ${\rm Tr} M = {\rm Tr} [U^\dag  M U]$,   invariant of the trace of  a matrix (or operator) under a unitary transformation.   Then, with a proper unitary transformation matrix,  both  the Green's function (or spectral function representation)  and velocity operator $V_{\alpha}$ are reduced to the  diagonal block matrices.  As a result,  the analytical form of the  real part of the conductivity of AANLG can be easily accessible. The details are as below.  First, by setting  $M=V_{\alpha}~ G(\omega+\Omega,{\bf k}) ~ V_{\beta}~  G(\omega,{\bf k}) $,  the trace of $M$ is ${\rm Tr} M = {\rm Tr} [U^\dag  V_{\alpha}~ G ~ V_{\beta}~  G U ]$. Then,  insert the identical matrix $UU^\dag  = I$ between the velocity  operator $ V_{\beta}$  and   Green's function   $ G$,  and the result $${\rm Tr} M  =  {\rm Tr}  [\mathcal{V}_{\alpha}~ \mathcal{G} ~ \mathcal{V}_{\beta}~  \mathcal{G} ]= {\rm Tr} \mathcal{M}  $$ is acquired. $\mathcal{G}= U^\dag  G  U  $  is the unitary transformation of   $G$ and it is related to the  spectral function representation $\mathcal{A}$ in such a manner:
 \begin{eqnarray}
 \mathcal{G}_{mn}(z)= \int^{\infty}_{-\infty} \frac{d \omega}{2\pi}
\frac{\mathcal{A}_{mn}(\omega')}{z-\omega'}.
\end{eqnarray}
 $ \mathcal{V}_{\alpha}=U^\dag   V_{\alpha}U $ is the unitary transformation of   $V_{\alpha} $.  Thirdly,  after the unitary transformation, both the $\mathcal{V}$ and   $\mathcal{A}$ are the diagonal block matrices.  That is to say,  the operator  $\mathcal{M}=\mathcal{V}_{\alpha}~ \mathcal{A} ~ \mathcal{V}_{\beta}~ \mathcal{A}$ is also  a block diagonal matrix, such as, $\mathcal{M}=\mathcal{M}_1\oplus \mathcal{M}_2 \oplus \mathcal{M}_3 \oplus \cdots$. Each $\mathcal{M}_j=\mathcal{V}_{j,\alpha}~ \mathcal{G}_j ~ \mathcal{V}_{j,\beta}~  \mathcal{G}_ j$ is a two by two matrix.  Finally,   the relation $Tr(\mathcal{M})=Tr(\mathcal{M}_1)+Tr(\mathcal{M}_2)+  Tr(\mathcal{M}_3)+\cdots$ is used to obtain  the AC conductivity for AANLG
\begin{subequations}
\begin{eqnarray}
\sigma_{\alpha \beta}(\Omega)
&=& \frac{e^2}{2\Omega}\int^{\infty}_{-\infty}\frac{d \omega}{2\pi}
[f(\omega-\mu)-f(\omega+\Omega-\mu)]
\times \int \frac{d^2 k}{(2\pi)^2}  [{\rm Tr} \mathcal{M}_1+{\rm Tr}  \mathcal{M}_2+ {\rm Tr} \mathcal{M}_3 + \cdots],\cr
&=& \sum^N_{j=1} \frac{e^2}{2\Omega}\int^{\infty}_{-\infty}\frac{d \omega}{2\pi}
[f(\omega-\mu)-f(\omega+\Omega-\mu)]
\times \int \frac{d^2 k}{(2\pi)^2}  {\rm Tr} \mathcal{M}_j,\\
&=& \sum^N _{j=1} \sigma_{j, \alpha \beta }(\Omega).
\end{eqnarray}
It is shown that the AC conductivity of an AANLG is equal to the summation of  the AC conductivity of each subsystem and $ \sigma_{j, \alpha \beta }(\Omega)$ of each  graphene-like layer  is
\begin{eqnarray}
\sigma_{j, \alpha \beta }(\Omega) = \frac{e^2}{2\Omega}\int^{\infty}_{-\infty}\frac{d \omega}{2\pi}
[f(\omega-\mu)-f(\omega+\Omega-\mu)] \cr
\times \int \frac{d^2 k}{(2\pi)^2} {\rm Tr} [ \mathcal {V}_{j,\alpha}~ \mathcal {A}_j(\omega+\Omega,{\bf k})
~  \mathcal{V}_{j,\beta}~   \mathcal {A}_j(\omega,{\bf k})].
\end{eqnarray}
\end{subequations}

The AC conductivity of  the AANLG  can be analytically specified. As the $2N \times 2N$ Hamiltonian is decomposed into $N$ $2 \times 2$ reduced Hamiltonian  matrices,  the effective Hamiltonian of each subsystem is described as Eq. (13).  Furthermore,  the Green's function,  spectral function representation and   velocity  operator associated with  each subsystem  are $2 \times 2$ matrices.   Thus, the analytical form of  AC conductivity of  each  graphene-like layer without ISOC is\cite{Tabert2012}
\begin{subequations}
 \begin{equation}
\begin{aligned}
\sigma_{j,xx }(\Omega) &=& \sigma_{intra} + \sigma_{inter},\quad{\rm without ~~ ISOC},\\
\end{aligned}
 \end{equation}
 \begin{equation}
 \begin{aligned}
 \sigma_{intra} &=& 4  \sigma_{0} \delta(\Omega)  |\mu- \varepsilon_j| \Theta (|\mu- \varepsilon_j|) ,\\
 \sigma_{inter} &=& \sigma_{0}~ \Theta (\Omega- 2| \varepsilon_j-\mu|),\\
  \label{Conduct-ISO}
 \end{aligned}
 \end{equation}
\end{subequations}
where $intra$ and $inter$  represent the contributions  resulting from the intraband and interband transitions, respectively.  With the ISOC  taken into consideration, the analytical form of  AC conductivity of each subsystem reads
\begin{subequations}
 \begin{equation}
\sigma_{j,xx }(\Omega) = \sigma(\Omega, | \varepsilon_j-\mu|),\quad{\rm with~~~~~ ISOC},\\
 \end{equation}
 where $\sigma(\Omega, | \varepsilon_j-\mu|)$ is  the conductivity for massive Dirac particles\cite{Gusynin2006,Tse2010}, and it is expressed as
 \begin{equation}
\frac{ \sigma(\Omega, \Upsilon)}{ \sigma_ 0} =  4 \frac{ \Upsilon^2-\Delta^2} { \Upsilon} \delta(\Omega)
 \Theta(\Upsilon-\Delta) + \Big[  1+ \Big(  \frac{2\Delta}{\Omega} \Big)^2 \Big]\Theta[(\Omega-2 {\rm max} (\Upsilon,\Delta))].
 \label{Conductivity-ISOC}
  \end{equation}
\end{subequations}
The dependence of  $\sigma_{j,xx }(\Omega) $ on the chemical potential $\mu$,   $\varepsilon_{j}$ (or $\varepsilon_{\bot}$), and  strength of ISOC  is  clearly revealed  through the afore-presented  formula.

The numerically calculated conductivity $\sigma_{xx }(\Omega) $  and the associated  conductivity of each  subsystem $\sigma_{j,xx }(\Omega) $ (denoted as sub-conductivity) of the AA-stacking trilayer graphene (TLG) are presented in Fig. 4.  Both the intraband and interband transitions contribute to  AC conductivity.  A delta peak at frequency $\Omega=0$, caused by the intraband transition,  is not shown here; that is, only the  conductivity resulting from the interband transitions  is shown.  The conductivity $\sigma_{1,xx }(\Omega)$, $\sigma_{2,xx }(\Omega)$ and $\sigma_{3,xx }(\Omega)$ of the subsystems are illustrated in the dashed curves. AC conductivity of TLG are presented in the solid curve, which is $\sigma_{xx}(\Omega)= \sum^{N=3}_{J=1} \sigma_{1,j }(\Omega) $, the superposition of  the AC conductivity of subsystems.  According  to  Eqs.  (\ref{Conduct-ISO}) and  (\ref{Conductivity-ISOC}), the profile of  each sub-conductivity $ \sigma_{j,xx }(\Omega)=\sigma_{0}~ \Theta (\Omega- 2|\varepsilon_j-\mu|)$   is governed by the step function  $\Theta(|\varepsilon_j-\mu|)$.  $\varepsilon_j$ related to TLG are  $\varepsilon_1= -\sqrt{2 \alpha^2_1 + V^2}$, $\varepsilon_2=0$, or $\varepsilon_3= \sqrt{2 \alpha^2_1 + V^2}$. In the absence of the gated potential ($V =0$) and  at $\mu=0$,  both  $\sigma_{1,xx } $  and $\sigma_{3,xx }$ show the absorption edge at $\Omega/ \alpha_1= \sqrt{2} $. $\sigma_{2,xx }(\Omega)$  contributes a constant background conductivity, which is  equal to $ \sigma_0$  (dashed curves in Fig. 4(a)). AC conductivity of TLG  (solid curve Fig. 4(a)) at high frequency is equal to a constant value, three times of  $\sigma_0$ .  At  $\mu=0.1 \alpha_1$,   $\sigma_{1,xx} $,  $\sigma_{2,xx}$  and  $\sigma_{3,xx} $ in Fig. 3(b))  show  step edges at frequencies $\Omega/ \alpha_1= \sqrt{2}+0.1 $,  $0.2$ and  $ \sqrt{2}+0.1 $. As a result, there  are three steps in the AC conductivity  (solid curve in Fig. 4(b)). In the application of the gated potential $V=0.4 \alpha_1$, the absorption edges of $\sigma_{1,xx } $  and $\sigma_{3,xx }$ occur at $\Omega= (\sqrt{2.16}+0.1)\alpha_1$ and $\Omega=(\sqrt{2.16}- 0.1)\alpha_1$ (Fig. 4 (c)).  The intrinsic SOC ($\Delta=0.1\alpha_1$) enhances the strength of sub-conductivity $\sigma_{2,xx}$ in the region $ 0.2 < \Omega /\alpha_1 < 0.6$. For comparison, the solid curves in Figs. 4(a)-4(d) are plotted in Fig. 4(e).  The characteristics of  $\sigma_{xx }(\Omega) $ of  TLG  are dependent on  $\mu$,  $V$, and  strength of ISOC.

The alternation of layer number $N$ has a great influence on  AC conductivity of an AANLG.
Figure 5 displays  $\sigma_{xx }(\Omega) $  and the associated  $\sigma_{j,xx }(\Omega) $  of the AA-stacking quad -layer graphene (QLG) .  At high frequency, AC conductivity  of QLG illustrates a constant value, which  is  equal to four times of $\sigma_0$.  There are two steps in the AC conductivity  of QLG  at $V=0$ (the solid blue curve in Fig. 5(a)).  The location of the absorption edge of  each   sub-conductivity  $\sigma_{j,xx }(\Omega)$ is controlled by the step function  $\Theta(|\varepsilon_j-\mu|)$.  The gated-potential-dependent   energy dispersions $\varepsilon_{j}$ related to  QLG are $\varepsilon_{\pm\pm}=\pm \sqrt{\texttt{B} \pm \sqrt{\texttt{B} ^2 - \texttt{C}}}$, where $\texttt{B}=\frac{3\alpha_1^2}{2}+\frac{5V^2}{4}$ and $\texttt{C}=\alpha_1^4 +\frac{3}{4}\alpha_1^2 V^2 +\frac{9}{16} V^4$.  In the absence of  the gated potential ($V=0$),   $\varepsilon_{++}=- \varepsilon_{-+}=2.618 \alpha_1$ and  $\varepsilon_{+-}= -\varepsilon_{--}=0.382\alpha_1$. The first and  second  absorption edges appear at $ \Omega /\alpha_1= 0.76$ and $ \Omega /\alpha_1= 5.2$ (Fig. 5(a)).  $\sigma_{1,xx }(\Omega)$ ($\sigma_{3,xx }(\Omega)$) is identical to $\sigma_{2,xx }(\Omega)$ ($\sigma_{4,xx }(\Omega)$).  At $\mu= 0.3 \alpha_1$, absorption edges occur at  $ \Omega /\alpha_1\approx0.2$,  $1.4$, $4.6$, and $5.8$.  AC conductivity features four steps (the solid cyan curve in Fig. 5(b)). The gated potential $V=0.3 \alpha_1$ modifies  $\varepsilon_{++}=2.881 \alpha_1$ and $\varepsilon_{+-}=0.344 \alpha_1$ and changes the locations of step edges (Fig. 5(c)).  The intrinsic SOC $\Delta=0.1 \alpha_1$   enhances the weight of AC conductivity around  $ \Omega /\alpha_1\approx0.2$. The solid curves in Figs. 5(a)-5(d) are plotted in Fig. 5(e) to illustrate that the effects  caused by the alteration of  $N$,  $\mu$,  $V$, and  strength of ISOC on  $\sigma_{xx }(\Omega) $ of an AANLG  are easily and clearly revealed through the analytical  formula.

\section {Conclusions}

In this  work, we propose an analytical model to derive the exact energy spectrum   and  dynamical conductivity in an AANLG in the presence of a bias voltage and spin-orbital coupling  at the same footing.
First,  a proper  transformation  matrix is built and used  to transform the $2N \times 2N$  tight-binding Hamiltonian matrix of  an AANLG into $N$ $2\times 2$ diagonal block matrices. Then,  an AANLG is reduced to $N$  graphene-like layers.  Thus, the exact energy spectrum  of a graphene-like layer is $E= \varepsilon _{\bot}\pm \varepsilon_{||}$.  $\varepsilon _{\bot}$, the effective on-site energy of the graphene-like layer, is controlled by  the interlayer interaction,  gated potential, and  layer number.  $ \varepsilon_{||}$ is the energy spectrum of a monolayer graphene  with SOC.  Furthermore, we  analytically study the dynamical conductivity of an AANLG, which is shown  to be the sum of the dynamical conductivity of $N$   graphene-like layers with/without SOC.  The dependence of  the dynamical conductivity  of  each graphene-like layer on the chemical potential,   $\varepsilon _{\bot}$, and the strength of SOC is clearly demonstrated.   Above all,  our model can efficiently and exactly acquire  the energy spectrum  and  dynamical conductivity in a gated AANLG with SOC.

\section*{Acknowledgements}
The author gratefully acknowledges the support of the Taiwan National Science Council under the Contract Nos. NSC 102-2112-M-165-001-MY3.\\
\bibliographystyle{rsc}
\bibliography{rsc}

\newpage
{\Large\bf Figure Captions}\\
\begin{itemize}
\item[FIG. 1.] The geometric structure of the AA-stacking multilayer graphene and the intralayer and interlayer interactions.
\item[FIG. 2.] Calculated energy dispersions around the Dirac point $\textbf{K}$ of the  AA-stacking bilayer graphene for different gateed potential $V$, intrinsic SOC $\Delta$, and Rashba SOC $\lambda_R$. The dashed curves in the inset: $(V, \Delta, \lambda_R) = (0,0,0)$; red solid curves: $(0.4,0,0) \alpha_1$; blue solid curves: $(0.4,0.1,0) \alpha_1$ ; green solid curves: $(0.4,0.0,0.05) \alpha_1$.

\item[FIG. 3.] Calculated energy dispersions of  the  AA-stacking trilayer graphene for different $V$, $\Delta$, and $\lambda_R$. The dashed curves in the inset: $(V, \Delta, \lambda_R) = (0,0,0)$; red solid curves: $(0.4,0,0) \alpha_1$; blue solid curves: $(0.4,0.1,0) \alpha_1$; green solid curves: $(0.4,0.1,0.05) \alpha_1$.

 \item[FIG. 4.]   AC conductivity $\sigma_{xx }$  (solid curves)  and  sub-conductivity  ($\sigma_{1}$, $\sigma_{2}$,  $\sigma_{3}$)   (dashed   curves) of the AA-stacking trilayer graphene for different $V$, $\Delta$, and $\mu$ are presented.  (a)  $(V,  \Delta,  \mu) = (0, 0, 0)$.  (b)  $(V, \Delta, \mu) = (0, 0, 0.1) \alpha_1 $.
    (c)  $(V, \Delta, \mu) = (0.4, 0, 0.1) \alpha_1 $.   (d)   $(V, \Delta, \mu) = (0.4, 0.1, 0.1) \alpha_1 $.
    The solid curves in (a)-(d) are plotted in (e).

\item[FIG. 5.]   Same plot as Fig.3 but for AA-stacking quad-layer graphene.   (a)  $(V,  \Delta,  \mu) = (0, 0, 0)$.  (b)  $(V, \Delta, \mu) = (0, 0, 0.3) \alpha_1 $.     (c)  $(V, \Delta, \mu) = (0.3, 0, 0.3) \alpha_1 $.   (d)   $(V, \Delta, \mu) = (0.3, 0.1, 0.3) \alpha_1 $.   The solid curves in (a)-(d) are displayed in (e).
\end{itemize}

\newpage
\pagestyle{empty}
\begin{figure}
\center
\includegraphics [width=0.9\textwidth, height=0.9\textheight]{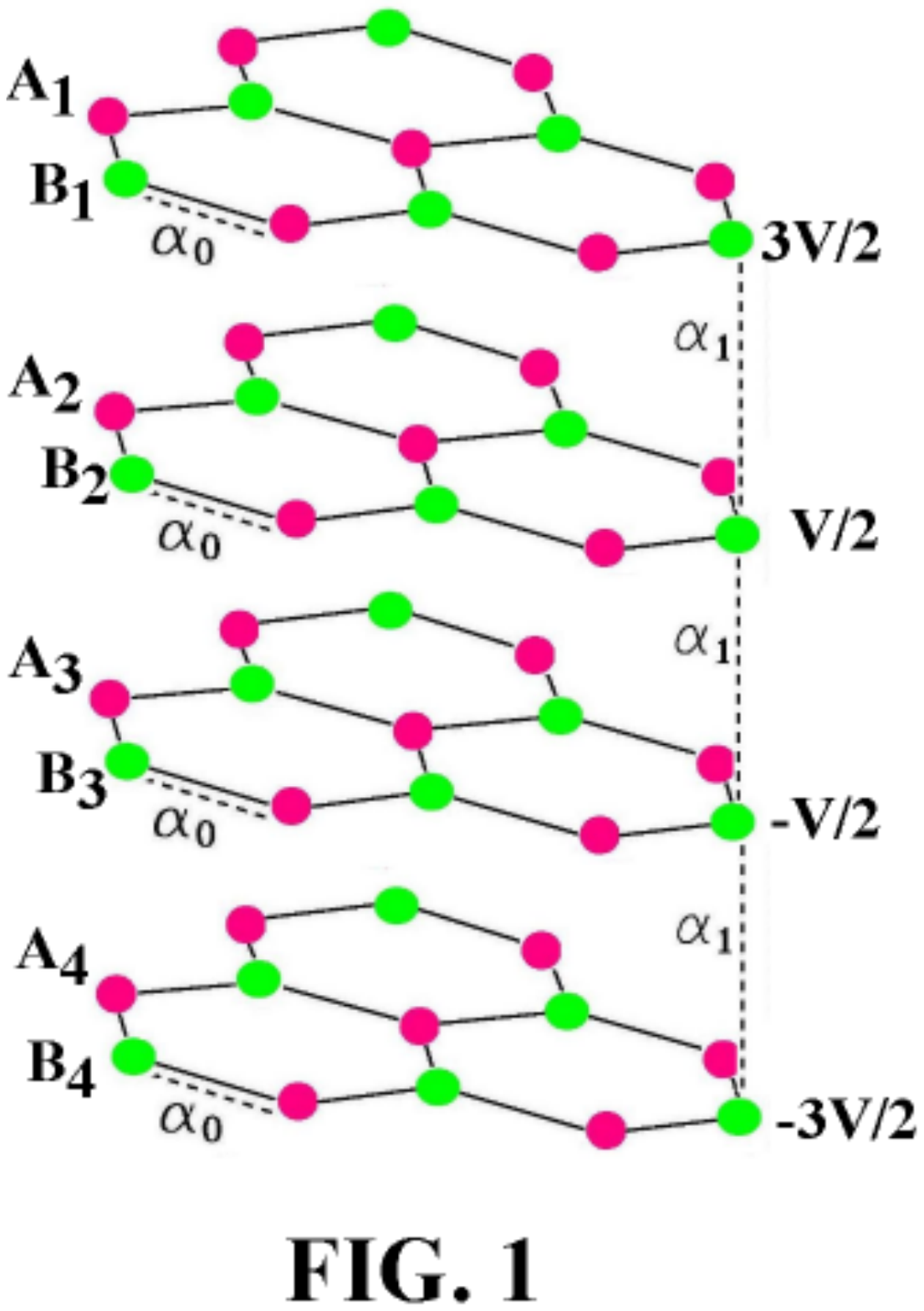}
\end{figure}

\newpage
\pagestyle{empty}
\begin{figure}
\center
\includegraphics [width=0.9\textwidth, height=0.9\textheight]{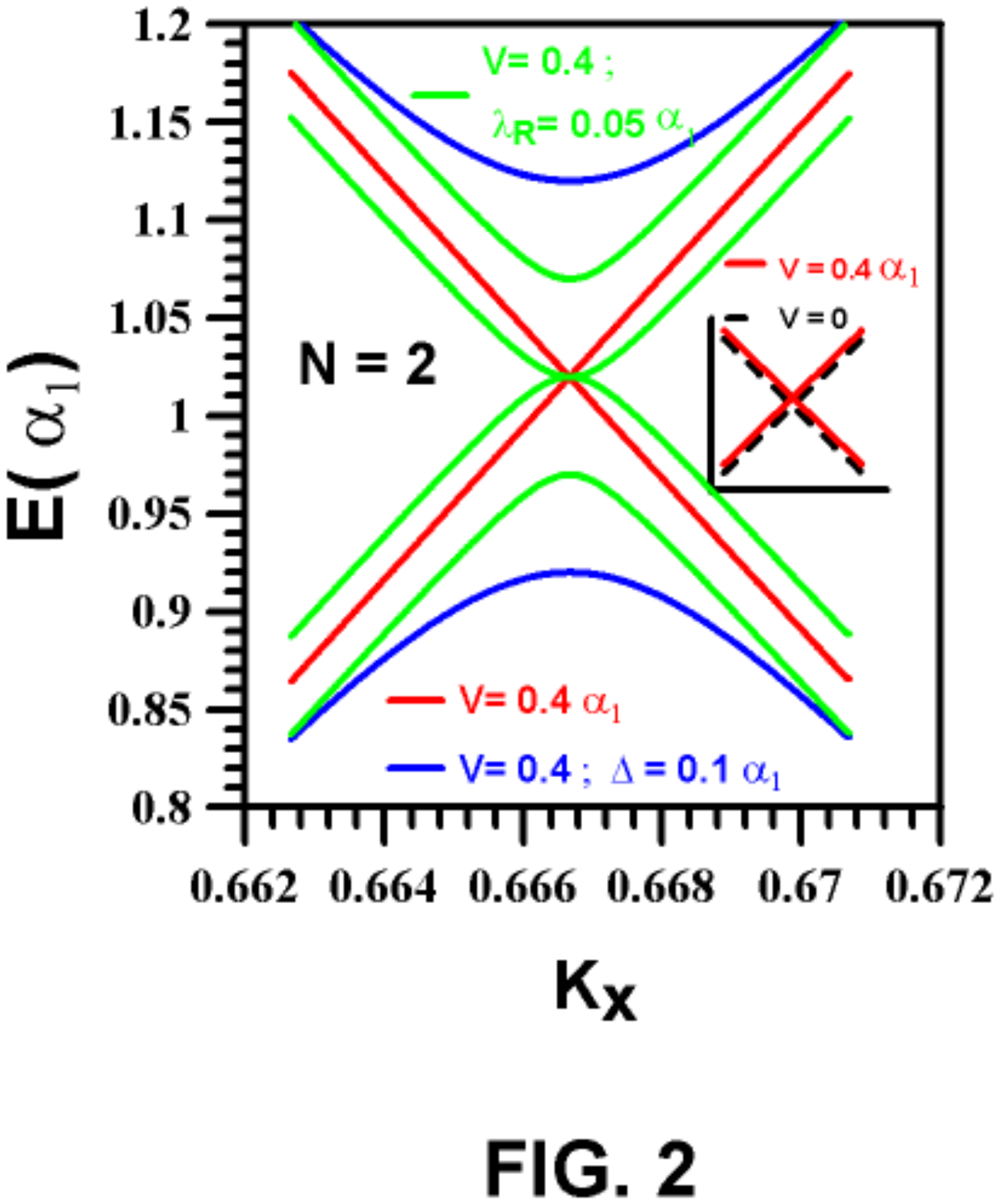}
\end{figure}

\newpage
\pagestyle{empty}
\begin{figure}
\center
\includegraphics [width=0.9\textwidth, height=0.9\textheight]{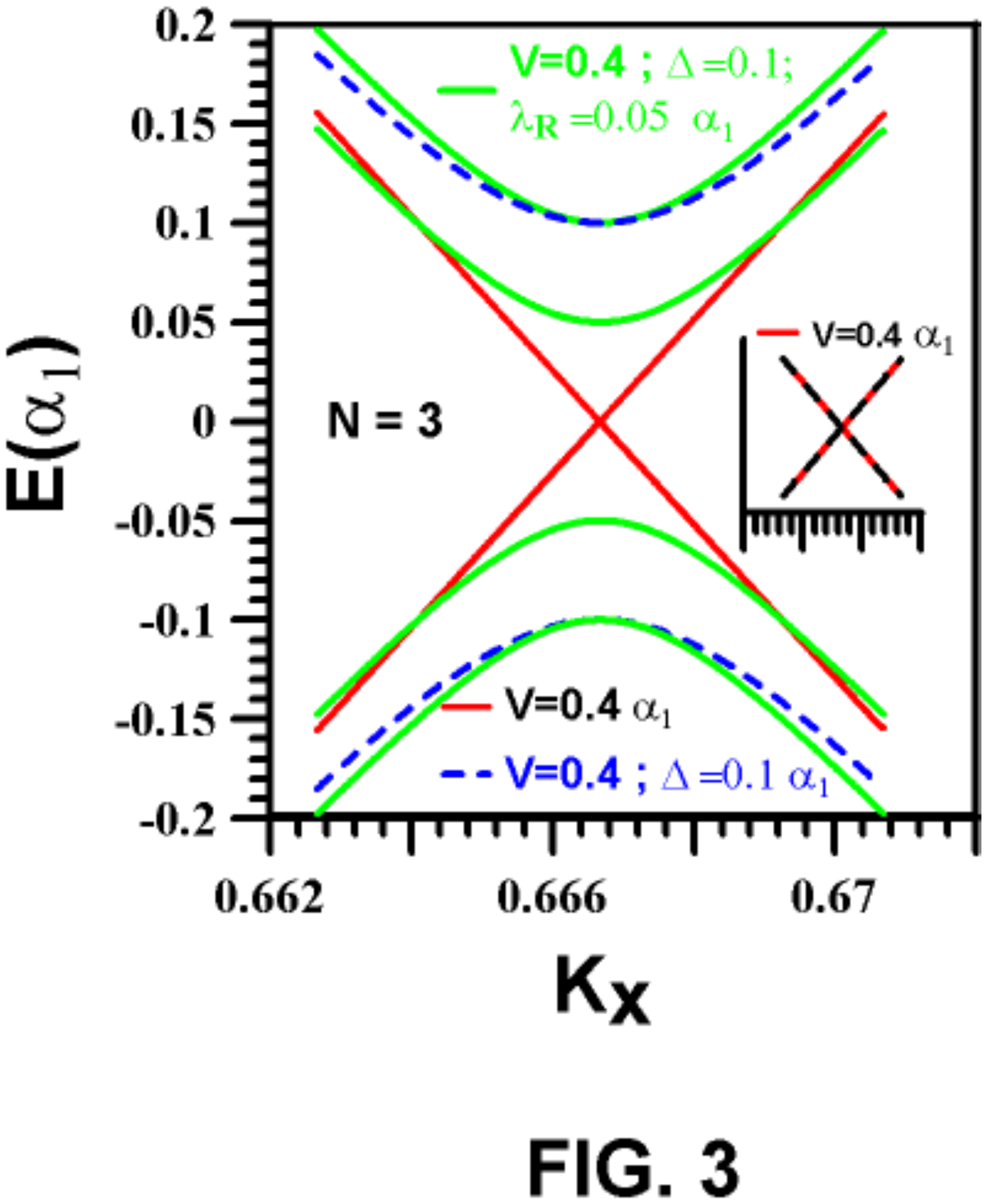}
\end{figure}

\newpage
\pagestyle{empty}
\begin{figure}
\center
\includegraphics [width=0.9\textwidth, height=0.9\textheight]{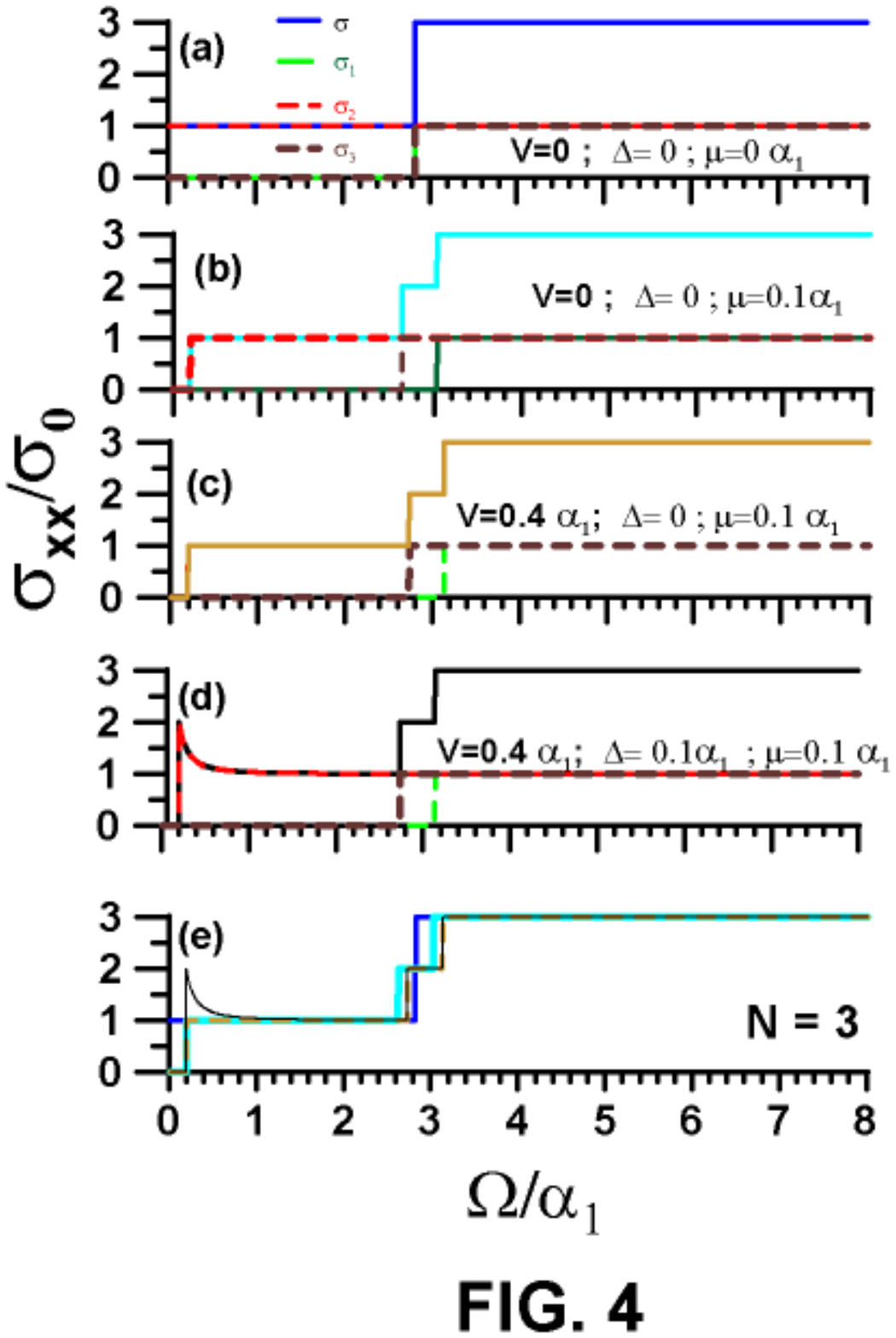}
\end{figure}

\newpage
\pagestyle{empty}
\begin{figure}
\center
\includegraphics [width=0.9\textwidth, height=0.9\textheight]{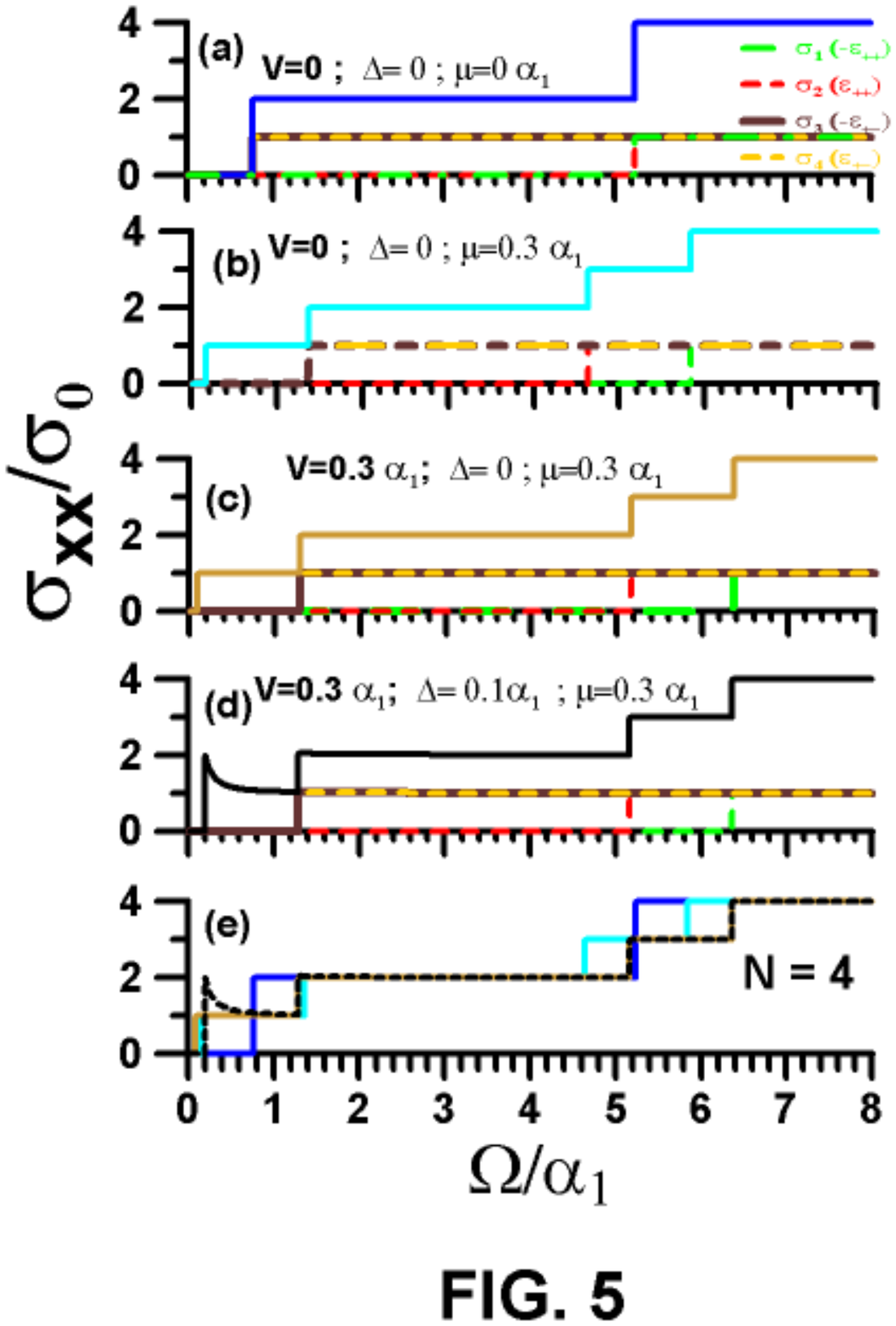}
\end{figure}

\end{document}